\documentclass[letterpaper]{jpconf}

\usepackage{graphicx}
\usepackage{cite}

\begin{document}

\title{Quantum Monte Carlo study of strongly interacting Fermi gases}
\author{Stefano Gandolfi}
\address{Theoretical Division, Los Alamos National Laboratory
Los Alamos, NM 87545, USA}
\ead{stefano@lanl.gov}

\begin{abstract}
In recent years Quantum Monte Carlo techniques provided to be a 
valuable tool to study 
strongly interacting Fermi gases at zero
temperature. We have used QMC methods to investigate several properties of the
two-components Fermi gas at unitarity and in the BCS-BEC crossover, both 
with equal and unequal masses corresponding to the $Li-K$ Fermi 
mixture.
In this paper we present several recent QMC results, including the energy
at zero and finite effective range,
the contact parameter and the static structure factor,
which, at low momentum,
depends strongly on the phonons in the unitary Fermi gas.
\end{abstract}

\section{Introduction}
\vspace{0.3cm}

During the last years important efforts have been devoted to study
ultra-cold Fermi gases, both experimentally and theoretically (for a review
see for example Ref.~\cite{Giorgini:2008}). By means of
Feshbach resonance it is possible experimentally to adjust the interaction
between atoms. The system can be tuned to a BCS state where Fermions
are weakly interacting, to the so called unitary regime where the two-body
scattering length is infinite, or to form a BEC condensate of Bosons.
Theoretically, ultra-cold Fermi gases have become an important
testing ground for a host of many-body methods, as well as an avenue to 
studying new physics.

The study of ultra-cold Fermi gases is very intriguing, and the unitary
limit became a very interesting many-body problem to solve for several
reasons. These systems can be experimentally studied at very low
temperatures of the order of 0.1 T$_F$, essentially zero temperature.
They are very dilute, and their properties are independent 
of the
form of the Hamiltonian used to describe the system, the results
given by any model satisfying the limits of large scattering length and
small effective range must be identical. Moreover these systems are very
strongly interacting, methods based on perturbation theory, mean-field,
or expansions of small parameters
are not obviously convergent, and the calculation of 
the properties of the Unitary Fermi gas is very challenging.  
The properties scale with powers of the Fermi momentum $k_F=(3\pi^2\rho)^{1/3}$.
The interparticle separation $r_0$  is
approximately given by $r_0=(9\pi/4)^{1/3}/k_F$, and the dilute regime
is guarantee by imposing $r_0 \gg r_{e}$, where $r_{e}$ is the
effective range of the two-body interaction. The universal regime is
given when $r_e \ll r_0 \ll a$, where $a$ is the scattering length, and
the unitary limit correspond to the case of $|a|=\infty$. In the limit
of $r_e=0$ the only scale of the system is $r_0$ or more conveniently the
Fermi momentum $k_F$.  Thus a natural unit for the energy if given by the
Fermi-gas energy $E_{FG}=3\hbar^2k_F^2/10\,m$, and there are no other free parameters. 
For this reason, these systems are called universal.

Ultra-cold Fermi gases can also be realized with population
imbalance, and the system can eventually exhibit different
phases~\cite{Zwierlein:2006,Partridge:2006} where superfluid and normal
phases may coexist or phase separate~\cite{Pilati:2008,Lobo:2006}, or
other intriguing more exotic phases like LOFF or $p$-wave superfluidity
could appear~\cite{Bulgac:2006,Bulgac:2008}.
In the last few years a new direction of research has been undertaken:
the realization of trapped two--component Fermi gases with the mixture
of Fermions with different masses, in particular the $^6Li-^{40}K$
mixture has been addressed~\cite{Whille:2008,Spiegelhalder:2010,Kohstall:2012}.
The properties of systems with mass imbalance can be very
different from the equal masses case; for example, for majority light
population the Chandrasekhar--Clogston limit is very small and close
to zero~\cite{Gezerlis:2009}. Then, at unitarity, the two--components
Fermi gas with different masses may exhibit very different properties with
respect to the equal mass case~\cite{Gubbels:2009,Baarsma:2010}, even in
the BCS limit~\cite{Baranov:2008}.

Several predictions for equal mass Fermi gases 
were made possible through
Quantum Monte Carlo techniques (QMC), later confirmed by
experiments~\cite{Carlson:2003,Astrakharchik:2004,Carlson:2005,Bulgac:2006b,Carlson:2008,
Pilati:2008,Gezerlis:2009,Gandolfi:2011,Carlson:2011}.
The Variational Monte Carlo (VMC) and the Diffusion Monte
Carlo (DMC) 
enable accurate calculations of
the ground state
of strongly interacting Fermions at unitarity, and to study its
properties~\cite{Gandolfi:2011}. In addition, the unbalanced two-component 
Fermi gases with equal masses do not exhibit a sign problem 
using the Auxiliary Field Monte Carlo (AFMC)~\cite{Carlson:2011}, 
and this permits a very accurate calculation 
to benchmark other many-body calculations.
Experiments are possible at very low
temperatures of the order of fractions of $T_F$, and then the zero
temperature properties can 
be directly obtained from experiments.

In this paper we will review several properties 
calculated using QMC methods, including the exact calculation of the energy,
the role of the effective range, the BCS-BEC crossover, and other properties
including  the contact parameter and the static structure factor.

\section{The model and QMC methods}
\vspace{0.3cm}
\label{sec:ham}

The AFMC performed on a lattice through the use of auxiliary fields does
not suffer of the Fermion sign problem and permits to calculate exactly
the energy of the unpolarized Fermi gas~\cite{Carlson:2011}. 
The AFMC method uses a BCS wave function as a trial wave function.
In this section
we describe the variational wave function and the DMC method in the 
continuum. 

In our study we model the ground state of the system in the continuum using the
Hamiltonian
\begin{equation}
H=\sum_{i=1}^{N_l}\frac{-\hbar^2}{2m_l}\nabla_i^2
+\sum_{j=1}^{N_h}\frac{-\hbar^2}{2 m_h}\nabla_j^2
+\sum_{i,j}v(r_{ij}) \,,
\end{equation}
where $N_l$ and $N_h$ is the number of particles with mass $m_l$
and $m_h$.  Since we want to model a system that is dilute, particles
interact only in $s$-wave, and the potential $v_{ij}$ is non-zero
only between particles with opposite spin state or with different mass.  Several forms
of the potential $v_{ij}$ has been explored, and provided to be
equivalent in the limit of small effective range~\cite{Forbes:2012}. In most of our studies
we have used the P\"oschl-Teller already employed in several previous QMC
calculations~\cite{Carlson:2003,Chang:2004,Carlson:2005,Carlson:2008,Gezerlis:2009,
Morris:2010,Gandolfi:2011,Forbes:2011,Forbes:2012}:
\begin{equation} 
v(r)=-v_0\frac{\hbar^2}{m_r}\ \frac{\mu^2}\ {\cosh^2(\mu r)} \,, 
\end{equation} 
where $m_r$ is the reduced mass, and the parameters
$\mu$ and $v_0$ tune respectively the effective range $r_e$ and the
scattering length $a$ of the interaction. For this potential, the unitary
limit corresponding to the zero-energy ground state between two particles
is with $v_0=1$ and $r_{e}=2/\mu$.  
Most of the results presented in this work
have been obtained by fixing $r_{e} k_F\approx 0.03$ but in several cases we
considered different values to check effects due to the effective range
of the potential. It is also possible to use other forms of the 
potential $v(r)$, and the results are universal in the limit of 
small effective range as we shall discuss in the next sections.

The QMC calculations are performed using the 
Jastrow-BCS  form of the wave function
of Refs.~\cite{Carlson:2003,Gandolfi:2011}: 
\begin{equation}
\Psi_v({\bf r}_1\dots {\bf r}_N)=
F({\bf r}_1\dots {\bf r}_N) \, \Phi_{\rm BCS} \,.
\end{equation}

The antisymmetric part $\Phi_{\rm BCS}$ is a particle-projected BCS
wave function including pairing correlations.  It is given by
\begin{equation}
\Phi_{\rm BCS} = {{\cal A}}[\phi(r_{11'}) \phi(r_{22'}) ... \phi(r_{nn'})] \,.
\end{equation}
The operator ${\cal A}$ 
is an anti-symmetrization operator,
the unprimed coordinates are for spin-up or heavy particles and the primed are
for spin-down or light particles, and $n=N/2$ for the unpolarized case. The pairing
function is parametrized as
\begin{eqnarray}
\label{eq:pairfn}
\phi({\bf r}) &=& \sum_{n} \alpha_{k_n^2} \exp [ i {\bf k}_n \cdot {\bf r}] +\beta(r) \,,
\nonumber\\
\beta(r)&=& \tilde{\beta}(r)+\tilde{\beta}(L-r)-2 \tilde{\beta}(L/2)~,
\nonumber\\
\tilde{\beta} (r) &=& [ 1 + c b  r ]\ [ 1 - \exp ( -d b r )] \frac{\exp ( - b r )}{dbr}~.
\end{eqnarray}
In the infinite volume limit $\phi({\bf r})$ is a function of one scalar variable,
the product  $k_F r$.
The simulations use periodic boundary conditions in a finite volume.
In this finite volume,
the function $\beta(r)$ has a range of $L/2$, $L$ is the size of the simulation box, the value of $c$ is
chosen such that $\beta(r)$ has zero slope at the origin,
and $b$, $d$ and $\alpha_{k_m^2}$ are variational parameters.  Note that if
$\beta(r)=0$, and $\alpha_{k_m^2}=0$ for $k > k_F$, $\Phi_{\rm BCS}=\Phi_{FG}$,
where the latter is the wave function describing the non interacting
Fermi gas in the normal phase. As $\Phi_{FG}$ and $\Phi_{\rm BCS}$
are orthogonal, we can study the Fermi gases both in the superfluid
and in the normal phase~\cite{Lobo:2006,Pilati:2008,Gezerlis:2009}. The pairing wave
function used here contains several free parameters that have been optimized
by minimizing the energy of the system using VMC and following the
strategy of Ref.~\cite{Sorella:2001}.  In the 
BEC side of the transition,
when $1/a\,k_F\geq 2$, we found that using the pairing function as described in
Ref.~\cite{Astrakharchik:2004}, i.e. the two--body wave function instead
of the function of Eq. \ref{eq:pairfn}, gives lower energies.  Instead,
in the BCS case for $1/a\,k_F\leq -1$, the BCS wave function $\Phi_{BCS}$
gives almost the same energy of $\Phi_{FG}$ as the pairing becomes less
important to the equation of state.
The polarized systems can be simulated by extending the wave
function to include single-particle states for the unpaired
particles~\cite{Carlson:2003,Chang:2004}. Note that the wave function
previously described cannot reproduce more exotic phases like LOFF or
$p$-wave pairing.

The Jastrow $F(R)$ includes a short- and a long-range part. The short-range
term is given by
\begin{equation}
F_{sr}(R)=\prod_{i<j}f_{sr}(r_{ij}) \,,
\end{equation}
where the function $f_{sr}(r)$ acts only between particles with different spin or mass,
and it is obtained by solving the equation
\begin{equation}
\label{eq:jas}
-\frac{\hbar^2}{2m_r}\nabla^2f_{sr}(r)+\alpha v(r)f_{sr}(r)=\lambda f_{sr}(r) \,.
\end{equation}
The parameter lambda is obtained by imposing the boundary condition
$f(r>d_0)=1$, where the healing distance $d_0$ is a variational parameter,
and the quenching $\alpha$ is adjusted to be less than one only in the BEC regime.
The correct boundary conditions to the wave function are guaranteed
by constraining $d_0\le L/2$ where $L$ is the size of the simulation
box, but we found that a typical good choice is given by taking $d_0\approx L/10$.
By solving Eq.~\ref{eq:jas} we also assure the correct behavior of the
wave function at small distances so that the Jastrow function $f$ is
defined to have $\partial f/\partial r$=0 at the origin.  
The calculation in the deep BEC can be also improved by adding a short-range 
function $f_{sr}(r)$ between particles with same spin or mass, and solving the
equation above with a repulsive interaction.

In order to precisely calculate the static structure function at low 
momenta, it is important to include the effect of phonons as long-distance
correlations. In our calculation, we include a term 
\begin{equation}
F_{lr}(R)=\prod_{i<j}\exp\left[\gamma\sum_n\frac{\exp(-\beta|{\bf q_n}|)}{|{\bf q_n}|}
\exp(-i{\bf q_n}\cdot{\bf r_{ij}})\right] \,,
\end{equation}
where $\gamma$ and $\beta$ are new variational parameters~\cite{Vitali:2011}.
The VMC algorithm is then used to optimize over wave functions
of this form. Diffusion Monte Carlo (DMC) is then used to project out the
ground state from:
\begin{equation}
\Phi_0=\lim_{\tau\rightarrow\infty}\Psi(\tau) \,,
\end{equation}
where
\begin{equation}
\Psi(\tau)=e^{-(H-E_T)\tau}\Psi(0)=e^{-(H-E_T)\tau}\Psi_v \,.
\end{equation}
The factor $E_T$ is a constant used to control the normalization of the
ground state.
The propagation is achieved using a many-body propagator defined as
\begin{equation}
G(R^\prime,R,\delta\tau)=\langle R^\prime\vert e^{-H\delta\tau}\vert R\rangle \,,
\end{equation}
and
\begin{equation}
\Psi(R^\prime,\tau+\delta\tau)=\int dR G(R^\prime,R,\delta\tau)\Psi(R,\tau) \,.
\end{equation}

A good approximation of $G$ can be obtained using a Trotter expansion, and
iterating the above equation for small values of $\delta\tau$.
Note that for a zero-range interaction the exact two-body propagator could also
be used to sample the many-body Green's function~\cite{Carlson:2012}.
In the case of Fermions there is a sign problem that needs to be taken care. 
A common approach combined with the DMC method is the fixed-node approximation.
The sampling of paths is restricted to regions where the trial wave function
is positive, and the problem is then recasted into a Bosonic problem (without
sign problem) in a restricted subspace. This approximation results in providing
an upper bound to the exact energy of the system, whose accuracy depends to the
quality of the variational wave function. 
More details on the implementation of the VMC and DMC algorithms can be
found in Ref.~\cite{Foulkes:2001}.

\section{The BCS-BEC crossover and the unitary limit}
\vspace{0.3cm}

\begin{figure}
\begin{center}
\includegraphics[width=0.7\textwidth]{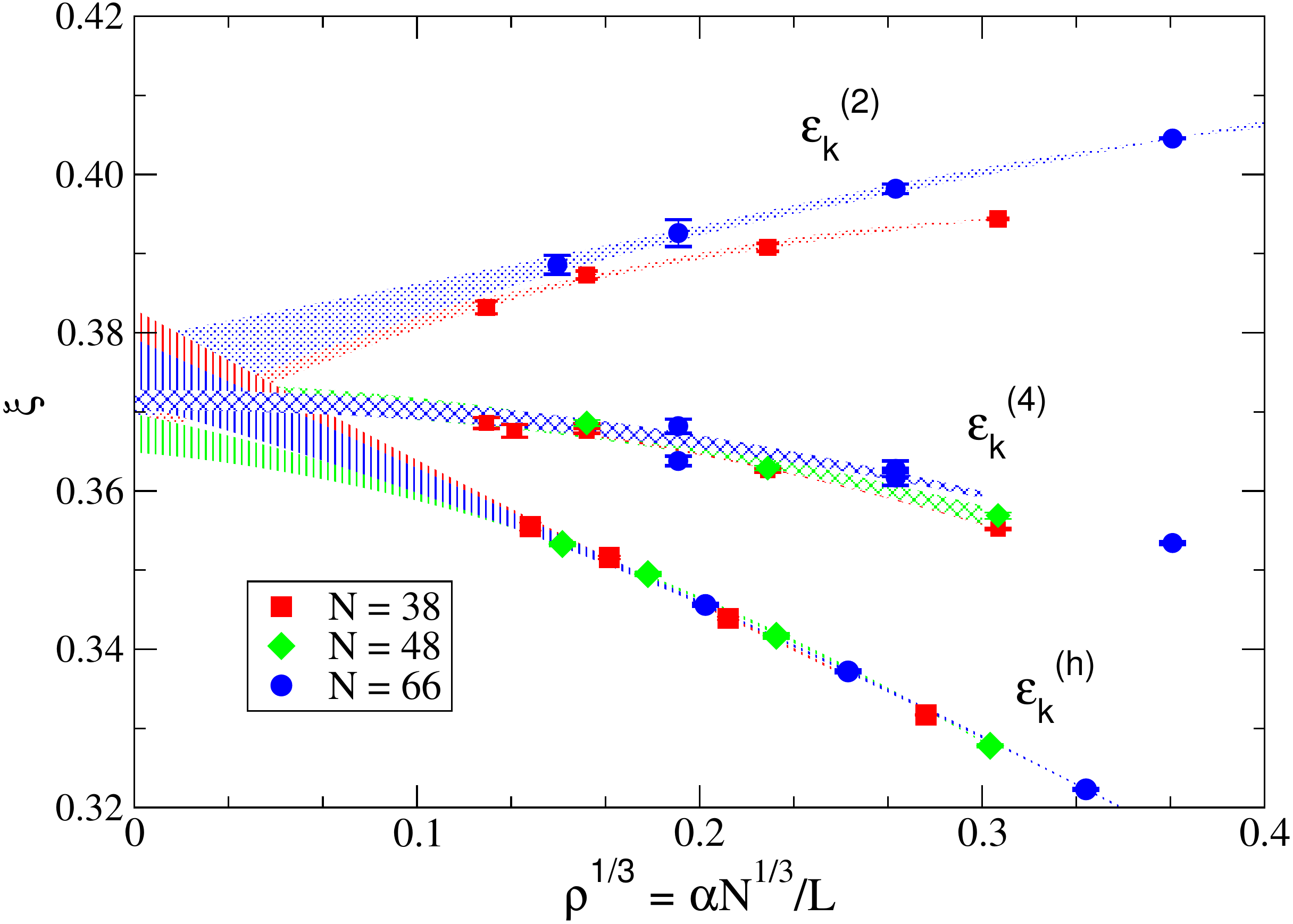}
\end{center}
\caption{Ground state energy of the unitary Fermi gas calculated using the AFMC method.
We show the value of $\xi$ as a function of the lattice size for various particle numbers $N$ 
and Hamiltonians, $\alpha$ is the lattice spacing. 
See Ref.~\cite{Carlson:2011} for more details.}
\label{fig:lattice}
\end{figure}

The BCS-BEC crossover has been calculated extensively 
over the past ten years.
The unitary limit is very interesting because perturbative methods cannot be
applied. In addition, there are no small parameters to expand, and thus using 
numerical calculations is the only way to accurately explore the properties of the system.
The energy of these systems is given by 
\begin{equation}
E=\xi E_{FG} \,,
\end{equation}
where $\xi$ is called Bertsch parameter.  There have been
several calculations of $\xi$ based on QMC techniques, and
recently we used the AFMC method to calculate exactly the energy
of the system~\cite{Carlson:2011}, the results are summarized in
Fig.~\ref{fig:lattice}. Since the lattice space provides a natural
regularization of the Hamiltonian, it is possible to test different
models of kinetic energy or dispersion relations. In the figure, we
show the results obtained using different regularization of the kinetic
energy. All the results correspond the case of infinite scattering
length, and in the case of $\epsilon_k^{(h)}$ and $\epsilon_k^{(2)}$,
the results are obtained at finite values of effective range and need to
be extrapolated.  Results obtained using different number of particles are
included in the figure.  We note that finite size effects are basically
absent, as using 38, 48 or 66 Fermions yields to the same energy.
This fact is because the interaction is very short-range, and because
pairing of the system is very strong.  Particles behave like Bosons,
and finite size effects associated to the kinetic energy vanish.
The absence of sizeable finite size effects for more than $\sim$40
particles has been also shown with a DMC calculation~\cite{Forbes:2011}.
By extrapolating the results to zero effective range gives the result
of $\xi=0.372(5)$~\cite{Carlson:2012} that is in excellent agreement
with a subsequent experiment performed at MIT~\cite{Ku:2012}.
The best upper bound estimate using the fixed-node DMC method is
$\xi=0.389(1)$~\cite{Forbes:2012}.

\begin{figure}
\begin{center}
\includegraphics[width=0.7\textwidth]{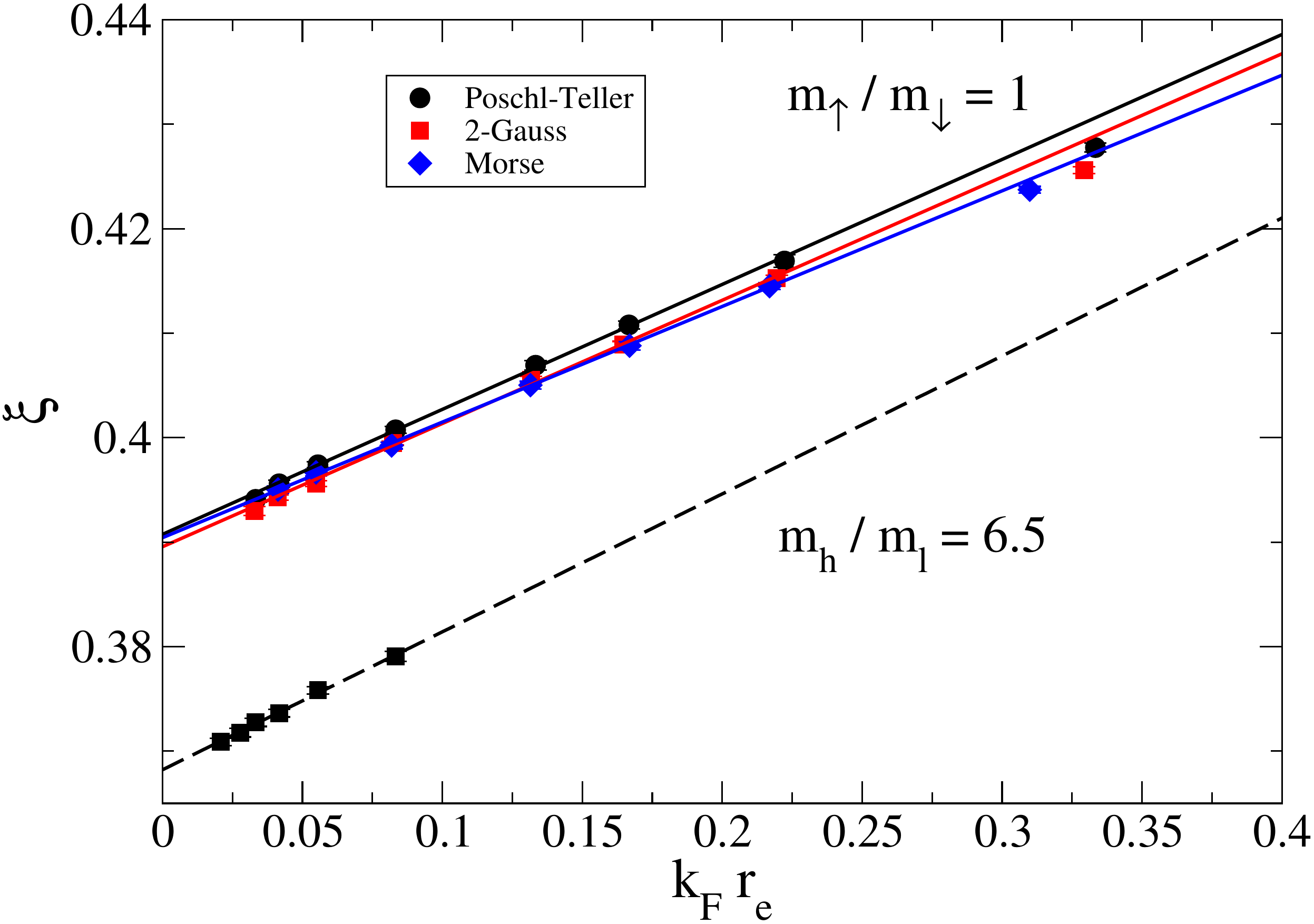}
\end{center}
\caption{The value of $\xi$ as a function of $k_F\,r_e$ calculated using DMC for 
the two-components with equal and unequal masses unitary Fermi gas.
In the case of equal masses, the results obtained using different two-body potentials
are presented. See Ref.~\cite{Forbes:2012} for more details.}
\label{fig:reff}
\end{figure}

Another quantity of interest is the dependence of the energy to the effective range:
\begin{equation}
\label{eq:reff}
\xi=\xi_0+s\,k_F r_e+\dots \,.
\end{equation}
At unitarity the parameter $s$ has been calculated using DMC and AFMC, and 
the different calculations yield $s=0.11(3)$ and $s=0.12(2)$~\cite{Carlson:2011,Forbes:2012}. 
In Fig.~\ref{fig:reff} we show the
results obtained using DMC with various potentials for the equal mass case, and 
for a mass ratio $m_h/m_l=6.5$ corresponding to the $Li-K$ mixture. 
It is pretty interesting that the value of $s$ does not depend
to the mass ratio of the two species.

\begin{figure}
\begin{center}
\includegraphics[width=0.8\textwidth]{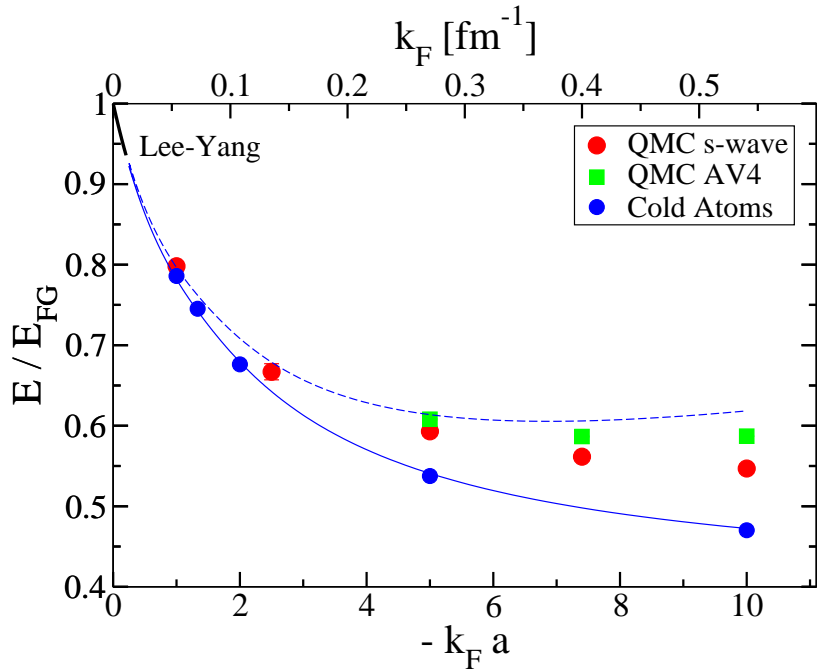}
\end{center}
\caption{The equation of state of cold atoms and low-density neutron matter. For the 
latter, we show results obtained from the $s$-wave part of a realistic neutron-neutron 
potential, and those obtained by extrapolating the cold-atoms results to large effective 
ranges. See the text for details. This figure is taken from Ref.~\cite{Carlson:2012}.}
\label{fig:nmat}
\end{figure}

The knowledge of the energy dependence to the effective range is important
to compare properties of neutron matter and cold atoms. At low-densities,
neutron matter has similar properties to cold atoms: the interaction is
mainly $s$-wave, the effective range is small, and the scattering length
is large~\cite{Gezerlis:2008}. In order to make qualitative comparisons between
cold atoms and neutron matter it is important to take into account the fact that
the neutron-neutron interaction has a small but finite $r_e$. Using the estimate of
$s$ as a function of $a\,k_F$ opens the possibility to compare these two different
systems. An example of this comparison is reported in Fig.~\ref{fig:nmat}.
In the figure we compare the cold atoms results corresponding
to $r_e=0$, the neutron matter equation of state calculated using the $s$-wave component
of realistic nuclear forces~\cite{Gezerlis:2008,Gezerlis:2010}, and the results obtained 
by extrapolating the cold-atoms results to the effective range of neutron-neutron potential
using Eq.~\ref{eq:reff}.

\begin{figure}
\begin{center}
\includegraphics[width=0.7\textwidth]{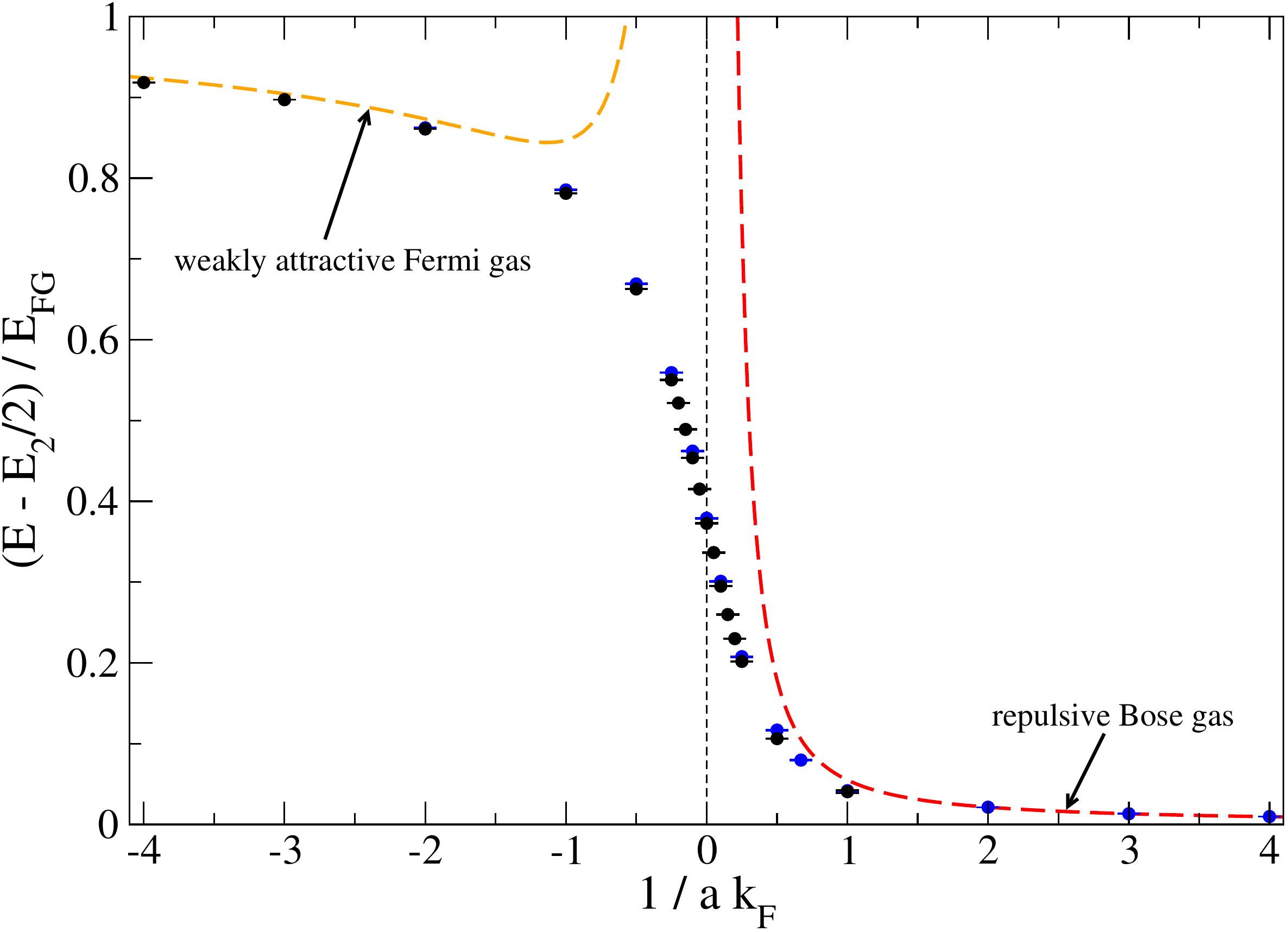}
\end{center}
\caption{The BCS-BEC crossover calculated using DMC for the two-component unequal
masses case corresponding to the $Li-K$ mixture. The two different set of points correspond
to the results obtained using different values of $r_e$. The dashed lines correspond to 
the perturbative calculation of weakly attractive Fermi gas in the BCS, and of the 
weakly repulsive Bose gas in the BEC.}
\label{fig:crossover}
\end{figure}

In the BCS-BEC crossover, the two limits correspond to the BCS where Fermions are weakly attractive
and to the BEC regime with very strong coupling. In the weak coupling
limit, the energy of the interacting Fermi gas is given by~\cite{Huang:1957}
\begin{equation}
\label{eq:bcs}
\frac{E}{E_{FG}}=1+\frac{10}{9\pi}a\,k_F+\frac{4(11-2\log 2)}{21\pi^2}(a\,k_F)^2+\dots \,,
\end{equation}
where $a$ is the scattering length and k$_F$ is the Fermi momentum of the system.
In the very strong coupling limit, when $a\,k_F\rightarrow 0^+$, the system
becomes made of Bosonic molecules, formed by one spin-up or heavy and one spin-down 
or light Fermion, that are weakly repulsive. Their energy is given by~\cite{Lee:1957}
\begin{equation}
\label{eq:bec}
\frac{E/N-E_2/2}{E_{FG}}=
\frac{10}{9\pi}a_{dd}k_F\frac{m_h m_l}{(m_h+m_l)^2}\times
\left[1+\frac{128}{15\sqrt{6\pi^3}}(a_{dd}k_F)^{3/2}+\dots\right] \,,
\end{equation}
where $a_{dd}$ is the Boson-Boson scattering length that can be obtained
from a few-body calculation~\cite{Petrov:2005}, or even by fitting QMC 
results~\cite{Astrakharchik:2004}.
The BCS-BEC crossover for a two-component $Li-K$ Fermi-Fermi mixture 
is shown in Fig.~\ref{fig:crossover}. In the figure, the blue and black points
correspond to the QMC results obtained using two different values of $r_e$ in the 
two-body interaction. 
In the unitary limit, the energy can be expanded as
\begin{equation}
\label{eq:eos}
\frac{E}{E_{FG}} = \xi-\frac{\zeta}{a\,k_F}-
\frac{5\nu}{3(a\,k_F)^2}+\dots \,.
\end{equation}
The fit to QMC results, extrapolated to $r_e=0$, gives $\xi=0.3726(6)$, $\zeta=0.900(2)$ and
$\nu=0.46(1)$.  The values of $\zeta$ and $\nu$ are very similar to the
case of equal masses reported in Ref.~\cite{Gandolfi:2011}.
In Fig.~\ref{fig:crossover} we also show the two limits given by
Eq.~\ref{eq:bcs} and~\ref{eq:bec} that agree with the QMC results.
The crossover corresponding to the $Li-K$ mixture is qualitatively similar
to the equal masses case~\cite{Astrakharchik:2004,Chang:2004}.

\section{Contact parameter and the static structure factor}
\vspace{0.3cm}

One of the most intriguing properties of strongly interacting Fermi
gases is the contact parameter. Shina Tan showed that several
quantities, including the pair distribution function, momentum
distribution, and the static structure factor in the limit of
small distances or high momenta are fully described by the contact
parameter~\cite{Tan:2008a,Tan:2008b,Tan:2008c}.  Also the equation of
state is related to the contact parameter.
This particular property is very interesting because the contact can
be calculated in different ways, and, more important, can be measured
in different experiments (see for example Refs.~\cite{Navon:2010,Kuhnle:2010,
Stewart:2010}).

In Ref.~\cite{Gandolfi:2011} several observables
related to the contact have been calculated using QMC methods. We note that, apart the fixed-node
approximation, any other observable depends to the choice of 
$\Psi_v$. Within DMC, 
we have computed observables with
\begin{equation}
\langle O \rangle=2 O_m - O_V \,,
\end{equation}
where
\begin{equation}
O_m = \langle \Phi_0 | O | \Psi_v \rangle \,,
\end{equation}
and
\begin{equation}
O_V = \langle \Psi_v | O | \Psi_v \rangle \,.
\end{equation}
Note that the energy is a special case, because the operator $H$ commutes with the 
propagator $G$, and in this case it's easy to see that $\langle H\rangle=H_m$.
The fact that several operators basically describe the same 
contact parameter is a proof of Tan's relations but also
of the very good quality of the trial wave function employed in
our calculations. 

\begin{figure}
\begin{center}
\includegraphics[width=0.7\textwidth]{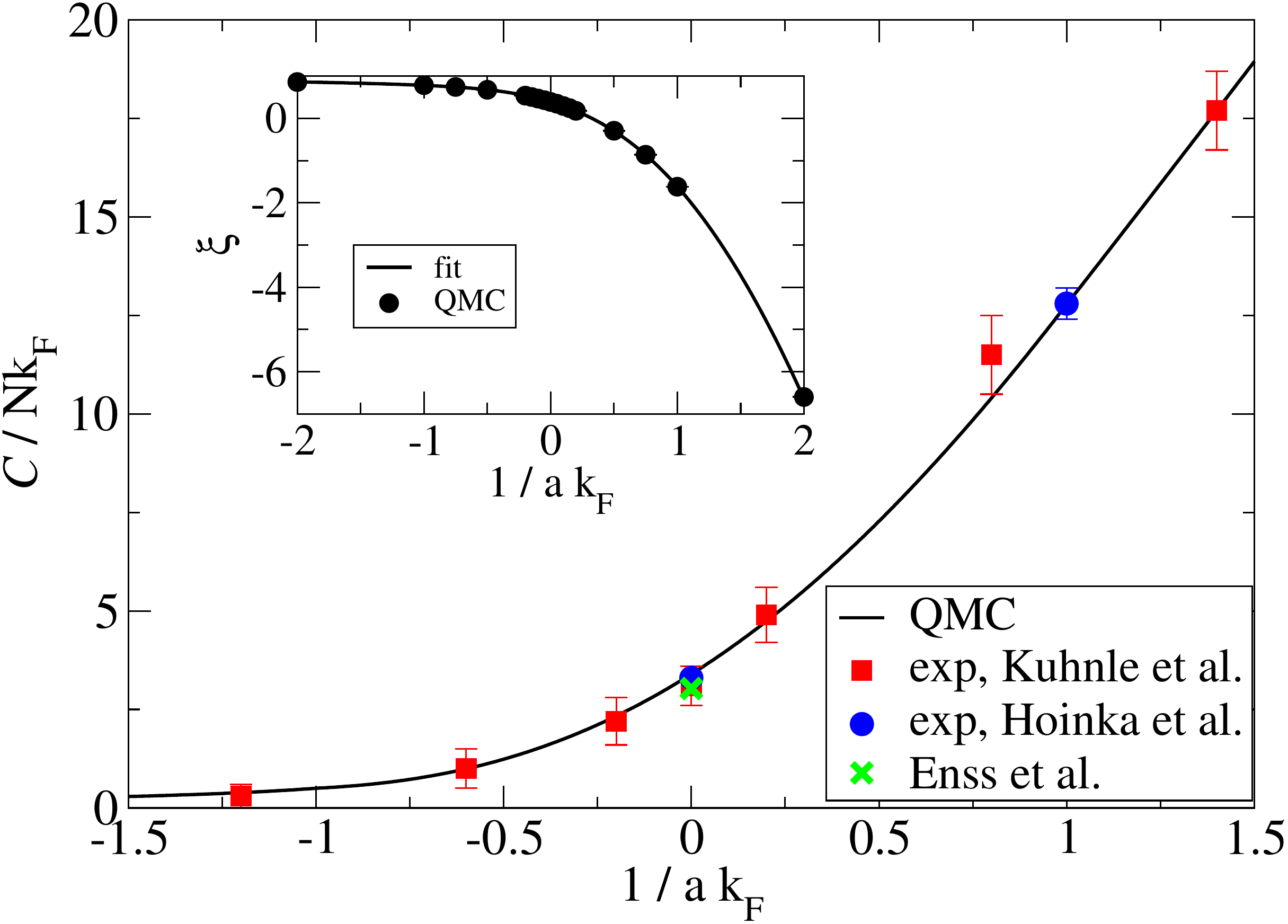}
\end{center}
\caption{The contact parameter as a function of $1/a\,k_F$ obtained from the
equation of state calculated using DMC (shown in the inset). The experimental 
points of Refs.~\cite{Hoinka:2013,Kuhnle:2011} and the result of Ref.~\cite{Enss:2011}
are also shown.}
\label{fig:contact}
\end{figure}

We have calculated the contact parameter in the BCS-BEC crossover 
using the adiabatic relation that relates the contact to the 
equation of state:
\begin{equation}
\frac{C}{Nk^{}_F}=-\frac{6\pi}{5}\frac{\partial \xi}{\partial (k^{}_F a)^{-1}} \,.
\end{equation}
The contact parameter is shown
in Fig.~\ref{fig:contact}, and compared with
recent measurements and with the result obtained by Enss 
\emph{et al.}~\cite{Enss:2011}.
In the inset we show the equation of state. The experimental
measurements based on the Bragg spectroscopy are in a very good agreement with
our calculations, both
at unitarity and in the BEC limit~\cite{Hoinka:2013}.

\begin{figure}
\begin{center}
\includegraphics[width=0.7\textwidth]{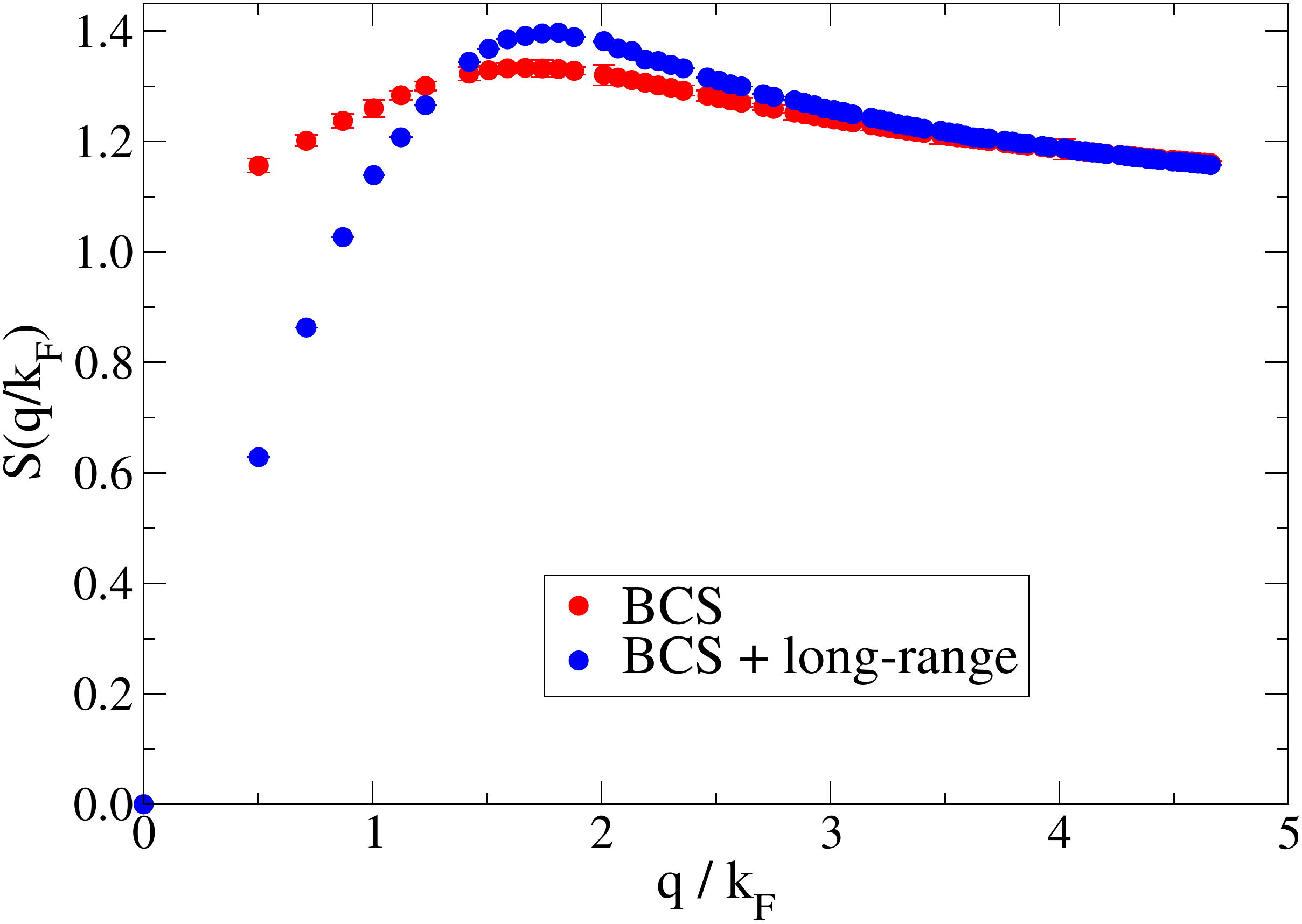}
\end{center}
\caption{The static structure factor (density-response sum rule) as a function of $q$.
The results have been obtained with the VMC, with and without long-range correlations in the Jastrow.
}
\label{fig:sq}
\end{figure}

One of the quantities related to the contact is the static structure factor $S(q)$,
that corresponds to the sum-rule of the density-response function:
\begin{equation}
S(q)=\int\,d\omega\,S(\omega,q) \,.
\end{equation}
The density-response function $S(\omega,q)$ has been measured by means
of the Bragg spectroscopy~\cite{Veeravalli:2008,Hoinka:2012,Hoinka:2013}.
At high momenta, this quantity is related to the contact through the relation
\begin{equation}
\frac{C}{Nk_F}=\frac{4q}{k_F}\left[\frac{S(q)-1}{1-4/(\pi\,q\,a)}\right].
\end{equation}
At low momenta $S(q)$ is related to the long-wavelength
phonons in the system.
The quantity $S(q)$ has been calculated using QMC for different values
of $a\,k_F$ in Ref.~\cite{Hoinka:2013}, and the agreement with experimental data
is within 1\%.
The calculation of $S(q)$ at small $q$ requires extra care. Due to
the phonon dispersion, we found it very important to include long-range
correlations in the variational wave function as discussed in
Sec.~\ref{sec:ham}.  In Fig~\ref{fig:sq} we show the VMC results for $S(q)$
obtained by including (or not) the long range correlations in the Jastrow.
We note that when the DMC algorithm is also used, the expectation
value of $S(q)$ is less dependent to the presence of long-range
correlations though even in this case the accuracy of the calculated
S(q) is much greater with the improved trial function. 
It would be very interesting to measure the behavior of $S(q)$ at small momenta,
where collective modes are expected to be important.

\section{Conclusions}
\vspace{0.3cm}

In this paper we present
recent results obtained using QMC methods
for two-component strongly interacting Fermi gases. 
The energy and its dependence on the product of the effective range 
and Fermi momentum have been
precisely determined theoretically and compared to experiment across
the BCS-BEC crossover. Other properties related to the contact have
also been
computed, in particular the static structure factor, and the results
are in very good agreement with experimental data.
Future directions include dynamic response of the unitary Fermi gas,
 inhomogeneous systems and the
transitions from three- to two-dimensions,  and multi-component Fermi gases.

\ack{ 
\vspace{0.3cm}
The author is grateful to  J. Carlson for very useful discussions and critical comments on
the manuscript.
The research presented in this paper has been supported by the U.S.~Department of Energy, Office of
Nuclear Physics, the NUCLEI SciDAC program and by the LANL LDRD program.
Computing time time were made available by Los Alamos Open Supercomputing.
This research used also resources of the National Energy Research
Scientific Computing Center, which is supported by the Office of
Science of the U.S. Department of Energy under Contract No.
DE-AC02-05CH11231.
}

\section*{References}
\vspace{0.3cm}

\bibliographystyle{iopart-num}

\providecommand{\newblock}{}

\end{document}